\pdfoutput=1
\documentclass[journal=aamick,manuscript=article]{achemso}

\usepackage{chemformula} 
\mciteErrorOnUnknownfalse
\usepackage[T1]{fontenc} 
\usepackage{siunitx}
\DeclareSIUnit\sccm{sccm}
\usepackage{indentfirst}
\usepackage{graphicx}
\usepackage{mathtools}

\usepackage{xpatch}
\makeatletter
\xpatchcmd\citenum{\NAT@spacechar}{\unkern}{}{\fail}
\makeatother
\author{Mattia Halter}
\affiliation{IBM Research GmbH - Zurich Research Laboratory, CH-8803 Rüschlikon, Switzerland}
\alsoaffiliation{Integrated Systems Laboratory, ETH Zurich, CH-8092 Zurich, Switzerland}
\email{att@zurich.ibm.com}

\author{Laura B\'egon-Lours}
\affiliation{IBM Research GmbH - Zurich Research Laboratory, CH-8803 Rüschlikon, Switzerland}

\author{Valeria Bragaglia}
\affiliation{IBM Research GmbH - Zurich Research Laboratory, CH-8803 Rüschlikon, Switzerland}

\author{Marilyne Sousa}
\affiliation{IBM Research GmbH - Zurich Research Laboratory, CH-8803 Rüschlikon, Switzerland}

\author{Bert Jan Offrein}
\affiliation{IBM Research GmbH - Zurich Research Laboratory, CH-8803 Rüschlikon, Switzerland}

\author{Stefan Abel}
\affiliation{IBM Research GmbH - Zurich Research Laboratory, CH-8803 Rüschlikon, Switzerland}

\author{Mathieu Luisier}
\affiliation{Integrated Systems Laboratory, ETH Zurich, CH-8092 Zurich, Switzerland}

\author{Jean Fompeyrine}
\affiliation{IBM Research GmbH - Zurich Research Laboratory, CH-8803 Rüschlikon, Switzerland}

\title[An \textsf{achemso} demo]
  {A back-end, CMOS compatible ferroelectric Field Effect Transistor for synaptic weights}


\begin{document}
%
%
%
%
%


\begin{abstract}

Neuromorphic computing architectures enable the dense co-location of memory and processing elements within a single circuit. 
This co-location removes the communication bottleneck of  transferring data between separate memory and computing units as in standard von Neuman architectures for data-critical applications including machine learning. 
The essential building blocks of neuromorphic systems are non-volatile synaptic elements such as memristors. Key memristor properties include a suitable non-volatile resistance range, continuous linear resistance modulation and symmetric switching.
In this work, we demonstrate voltage-controlled, symmetric and analog potentiation and depression of a ferroelectric Hf$_{0.57}$Zr$_{0.43}$O$_2$ (HZO) field effect transistor (FeFET) with good linearity. 
Our FeFET operates with a low writing energy (\si{\femto\joule}) and fast programming time (\SI{40}{\nano\second}). Retention measurements have been done over 4-bits depth with low noise (\SI{1}{\percent}) in the tungsten oxide (WO$_{\mathrm{x}}$) read out channel. By adjusting the channel thickness from 15nm to 8nm, the on/off ratio of the FeFET can be engineered from \SI{1}{\percent} to \SI{200}{\percent} with an on-resistance ideally \SI{>100}{\kilo\ohm}, depending on the channel geometry. The device concept is using earth-abundant materials, and is compatible with a back end of line (BEOL) integration into complementary metal-oxide-semiconductor (CMOS) processes. It has therefore a great potential for the fabrication of high density, large-scale integrated arrays of artificial analog synapses.
\end{abstract}

\section{Introduction}
The development of new computing architectures has seen a substantial push since the scaling of conventional CMOS technology has come to its limits and cannot keep up with the always increasing demand for computational power. A large part of today's computing resources is dedicated to processing large amounts of data, such as images, videos, or sensor outputs. 
For all these workloads, conventional von Neuman architectures are limited by a fundamental, time and power consuming task of transferring data between the processor and the memory.\cite{Wong2015} Brain-inspired neuromorphic architectures with co-located computation and memory units appear as promising candidates to overcome this issue.\cite{Poon2011} 
Such architectures consist of neurons that are interconnected by plastic synapses, which can be arranged in a crossbar topology to efficiently perform matrix-vector multiplications\cite{Gokmen2016} – a key computing task when executing neural networks\cite{Kim2017a,Yu2018}.
In recent years, much progress has been made in neuromorphic hardware, in particular in creating crossbar arrays of artificial synapses connected to CMOS neurons.\cite{Likharev2003,Nishitani2012,Mulaosmanovic2018,Burr2017a,Ha2011a,Chen2018b,Hansen2018} Multiple device concepts have been proposed in order to realize the required artificial synapse, such as phase change memory (PCM)\cite{Lacaita2006,Raoux2010a,Boybat2018}, filamentary and non-filamentary resistive switching memory (RRAM),\cite{Baek2004,Lee2008,Waser2009} electro-chemical memory (ECRAM)\cite{Fuller2017,Tang2018, Kim2019}, and ferroelectric (FE)\cite{Nishitani2012,Jerry2018,Aziz2018,Kim2019a,Mulaosmanovic2017} based memory cells.
Unlike classical memory elements, such devices are characterized by the stronger need for multilevel or analog programming capabilities to define the synaptic weight. 
While PCM and RRAM  devices are essentially current controlled, the different states in ferroelectric memory elements are controlled by applying an electric field. The states are linked to the partial switching of the ferroelectric polarization, which allows to fine tune the synaptic weights in analog computing approaches, with fast and low-power writing.\cite{Slesazeck2018a} 

For circuits solving real world applications, the number of required synapses rapidly explodes with the complexity of the task at hand. Solving even a simple task such as the MNIST database of handwritten digits requires \num{\approx e5} synapses\cite{Burr2015a}, while the training of a deep neural network (DNN) relies on up to millions of synapses. 
Such numbers of  hardware synapses can only be obtained in densely integrated circuits such as fabricated using modern CMOS technology. Part of the functions in neural networks can also be implemented using CMOS circuits (e.g. activation).  
Therefore,  it is important that materials and processes are CMOS compatible. The recent discovery of ferroelectricity in hafnia composites\cite{Boscke2011a}, a material already present in CMOS lines, has revived research activity in the field of integrated ferroelectrics. Artificial ferroelectric synapses have been realized based on two device concepts, namely two-terminal ferroelectric tunneling junctions (FTJ)\cite{Chanthbouala2012,Ambriz-Vargas2017a,Chen2018,Tian2017,Goh2018a} and three-terminal ferroelectric field-effect transistors (FeFET)\cite{Nishitani2012,Mulaos2017a,Sharma2017a,Krivokapic2017a,Kim2019a, Jerry2018,Mo2019}. 
Hafnia-based FTJs remain a challenge as the stabilization of the ferroelectric phase in sub-\SI{5}{\nano\metre} thick structures becomes difficult and polarization drops at film thicknesses relevant for tunneling.\cite{Tian2018,Goh2018a,Chernikova2016} 
Using a transistor instead has the advantage of separating the write process (low power write through high impedance gate\cite{Slesazeck2018a}) and the read process (through source-drain resistance). It also permits to tune synaptic resistance by changing the channel geometry. 
Hafnia-based FeFETs were demonstrated mainly as non-volatile memory cells\cite{Mulaos2017a}, steep-slope field-effect transistors\cite{Sharma2017a,Krivokapic2017a}, and artificial neurons\cite{Mulaosmanovic2018}. 
These concepts usually are implemented on the front end of line (FEOL) and use Si as a channel. Because of the constraints imposed by the FEOL on the thermal budget and on the device geometry, an integration in the back end of line (BEOL) can be advantageous. 
E.g., an integration in the BEOL enable a larger device area with respect to the size of the ferroelectric domains, which can translate into a larger number of states. 
Recently, analog synaptic behavior has been shown in a hafnia-based FeFET with indium gallium zinc oxide (IGZO) and poly-Si channels fabricated in the BEOL.\cite{Kim2019a,Jerry2018,Mo2019} 
The combination of a hafnia-based ferroelectric with an oxide channel is expected to alleviate the known issues associated with Si-based FeFETs such as unintended low-k interfacial layers formed at the Si interface. On Si based channels, buffer layers have been used as a solution, but they have the disadvantage of reducing the effective field over the ferroelectric layer.\cite{Kim2002,Sakai2004,Takahashi2005,Kaneko2011}
For neuromorphic applications the absolute resistance should be in the \si{\mega\ohm}range\cite{Gokmen2016}  and the relative change in resistance ideally within a window of  8\cite{Gokmen2016} up to 20-50\cite{Yu2015}. Those values are a compromise between being large enough for performing learning tasks, and low enough to avoid one synaptic element to dominate the respone of a whole column/row of the overall crossbar array.\cite{Gokmen2016,Yu2015}
Here, we report on a Hf$_{0.57}$Zr$_{0.43}$O$_2$ (HZO) based FeFET utilizing a tungsten oxide (WO$_{\mathrm{x}}$) channel. 
We demonstrate the impact of the ferroelectric polarization on the channel resistance, the influence of the channel thickness on the on/off ratio, ferroelectric HZO with a long endurance, the stabilization of multiple differentiable states, a good retention as well as a continuous potentiation and depression. By using a BEOL compatible process and by using only abundant and CMOS friendly materials, the proposed HZO/WO$_{\mathrm{x}}$ stack is very promising for large-scale integrated neuromorphic hardware based on ferroelectrics.

\section{Results and discussion}

\begin{figure*}[t]
	\includegraphics[width=\textwidth]{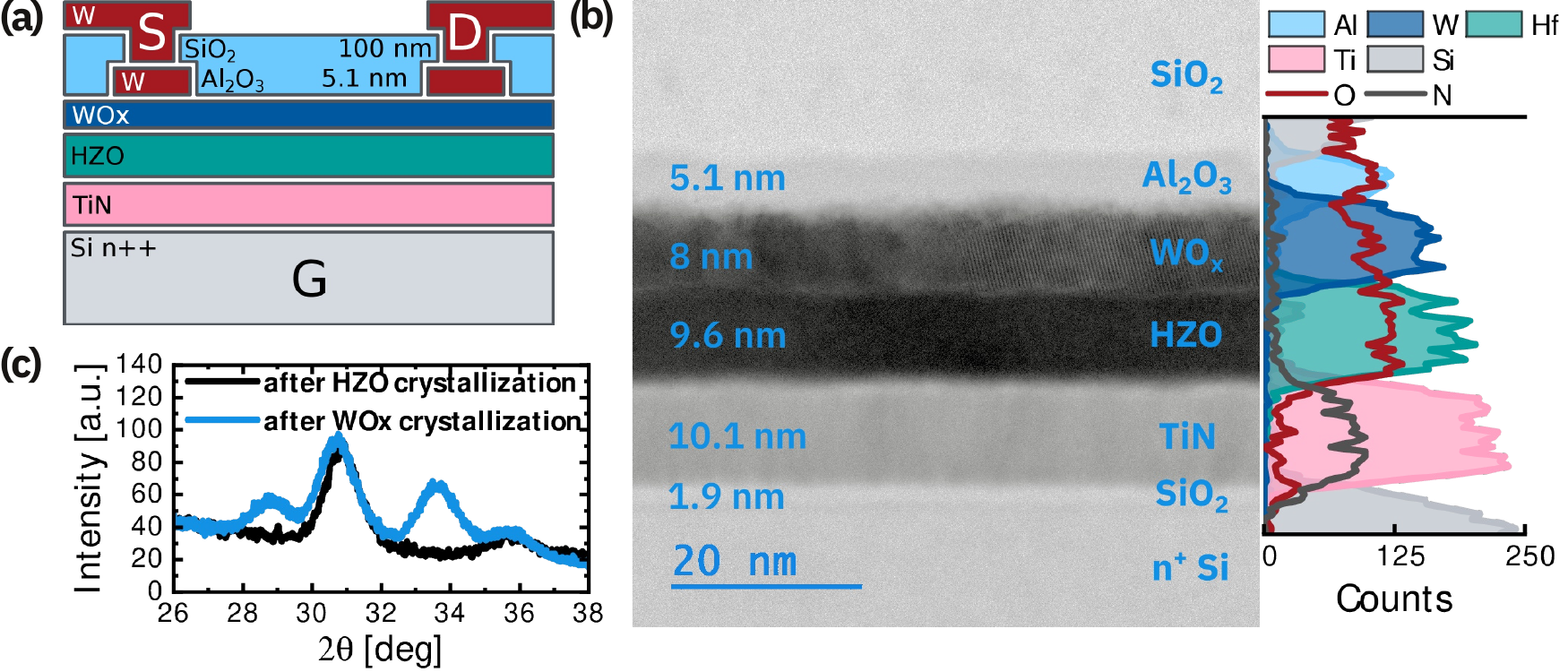}
  	\caption{Structural data of the FeFET. \textbf{(a)} Schematic illustration of a FeFET, indicating source (S), drain (D), gate (G), a WO$_{\mathrm{x}}$ channel and a ferroelectric HZO gate dielectric. \textbf{(b)} Cross-sectional BF-STEM image with energy-dispersive X-ray spectroscopy (EDS) line profile of the SiO$_2$/Al$_2$O$_3$/WO$_x$/HZO/TiN/$n^+$ Si gate region. \textbf{(c)} GIXRD for a diffraction angle (2$\theta$) from 26° to 38° showing the presence of the orthorhombic/tetragonal crystalline phase in HZO after crystallization and after the W layer was oxidized to WO$_{\mathrm{x}}$.}
  	\label{fig:structural}
\end{figure*}

For our study, we designed FeFET devices similar to back gated PseudoMOS\cite{Cristoloveanu2000} with an HZO (\SI{10}{\nano\metre})/TiN (\SI{10}{\nano\metre})/$n^+$ Si gate stack and an \SI{8}{\nano\metre} thick WO$_{\mathrm{x}}$ channel\cite{Tang2018, Kim2019} (Figure \ref{fig:structural}a). The channel is formed by oxidizing \SI{2.5}{\nano\metre} of W after the formation of the ferroelectric HZO\cite{OConnor2018}. The source and drain contacts are deposited on the WO$_{\mathrm{x}}$ channel through lift-off. The device is encapsulated by a \SI{5}{\nano\metre} Al$_2$O$_3$ and a \SI{100}{\nano\metre} SiO$_2$ passivation layer. Contact pads are formed on top of the passivation layers and routed through openings to source and drain. The gate is accessible through the highly $n^+$ doped Si substrate and is shared between all devices on our chip. 
As visible in the bright field scanning transmission electron microscopy (BF-STEM), our fabrication process results in sharp interfaces between the layers and crystalline WO$_{\mathrm{x}}$ grains (Figure \ref{fig:structural}b). The energy-dispersive X-ray spectroscopy (EDS) line profile confirms the targeted elemental distributions and reveals regions of intermixing between the various layers.
After the low temperature crystallization of HZO by a millisecond flash lamp technique described elsewhere\cite{OConnor2018}, grazing incidence X-ray diffraction (GIXRD) analysis shows the characteristic peak at 30.6° of the orthorhombic/tetragonal phase in HZO (Figure \ref{fig:structural}c). 
The diffractogram is consistent with data from metal-ferroelectric-metal (MFM) structures with the same HZO published in Ref. (\citenum{OConnor2018}). No monoclinic phase (peaks at 28.2° and 31.8°)\cite{Materlik2015} is present in our samples, which is a consequence of the low temperature crystallization technique. Following the oxidation and crystallization of W to WO$_{\mathrm{x}}$, GIXRD still shows no monoclinic HZO phase, but displays two additional peaks at 28.8° and 33.6° that can be attributed to the monoclinic P121/c1 phase of WO$_{\mathrm{x}}$ (ICSD-647640)\cite{Salje1975a}.

\begin{figure}[t!]
	\includegraphics[width=3in]{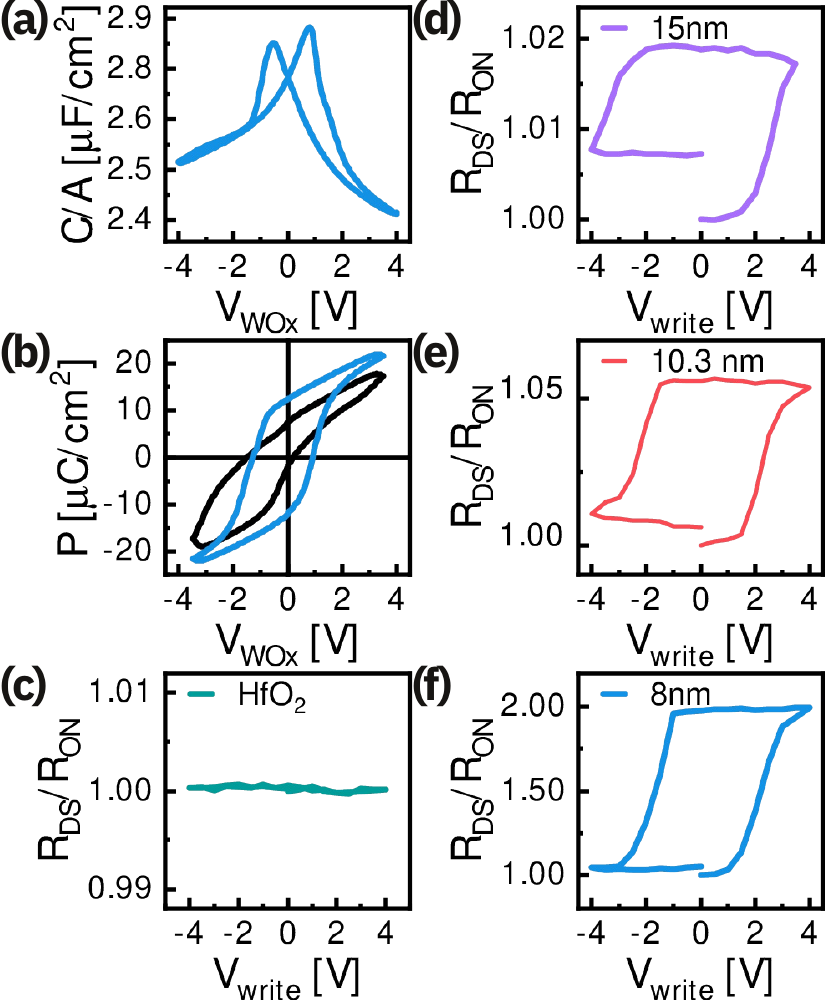}
	\captionsetup{width=3in}
  	\caption{Capacitance and polarization behavior of a \SI{60 x 60}{\micro\metre} W/WO$_{\mathrm{x}}$/HZO/TiN/n$^{\mathrm{+}}Si$ MSFM structure and WO$_{\mathrm{x}}$ channel resistance hysteresis: \textbf{(a)} Capacitance versus voltage ($C-V$) measurements after the HZO was woken up by 20 $C-V$ cycles.  \textbf{(b)} Polarization versus voltage ($P-V$) characteristics in the pristine state and after \num{e5} cycles. \textbf{(c, f)} Comparison of simultaneously processed samples with HZO and HfO$_2$ gate dielectric. The non ferroelectric HfO$_2$ sample does not show any channel resistance hysteresis. \textbf{(d, e, f)} Influence of the channel thickness ($d_{\mathrm{WOx}}$) on the on/off ratio.  }
  	\label{fig:WOx_thickness_HfO_PV_CV}
\end{figure}

For the electrical characterization of HZO in our FeFET devices, additional metal-semiconductor-ferroelectric-metal (MSFM) capacitor structures have been processed on the same sample. "Capacitance versus voltage" ($C-V$) measurements on a \SI{60 x 60}{\micro\metre} capacitor reveal a ferroelectric typical butterfly-shaped hysteresis curve typical of ferroelectrics, with a capacitance per unit area of $C_{OX}=\SI[per-mode=symbol]{2.7}{\micro\farad\per\centi\metre\squared}$ (Figure \ref{fig:WOx_thickness_HfO_PV_CV}a). The asymmetric behavior originates from the asymmetric electrodes (WO$_{\mathrm{x}}$, TiN). "Polarization versus voltage" ($P-V$) measurements were performed on the same capacitor (Figure \ref{fig:WOx_thickness_HfO_PV_CV}b) and show typical characteristics. 
In the pristine state, the $P-V$ curve is anti-ferroelectric (AFE)-like with hysteresis, especially on the negative voltage side.\cite{Kim2016} We applied \num{e5} switching cycles with an amplitude of \SI{\pm 3.8}{\volt} at a frequency of \SI{100}{\kilo\hertz}, resulting in a pinched $P-V$ curve with a positive (negative) remanent polarization $+P_{\mathrm{r}} = \SI[per-mode=symbol]{12.4}{\micro\coulomb\per\square\centi\metre}$ ($-P_{\mathrm{r}} = \SI[per-mode=symbol]{11.8}{\micro\coulomb\per\square\centi\metre}$). 
Furthermore, a slight imprint with a positive coercive voltage of $+V_{\mathrm{C}} = \SI{0.91}{\volt}$  and a negative one of $-V_{\mathrm{C}} = \SI{-1.27}{\volt}$  are observed due to the asymmetric electrodes. The cycling endurance of our HZO is \num{e8} for an MFM structure and \num{8e6} in the case of  the MSFM configuration present in our FeFET (Figure S1).

Having confirmed the ferroelectric nature of our HZO gate dielectric, the electrical characterization of the WO$_{\mathrm{x}}$ channel in a FeFET device was performed next, by investigating the influence of $P_{\mathrm{r}}$ , channel thickness ($d_{\mathrm{WOx}}$), and the channel carrier concentration ($N_D$) on the channel resistance ($R_{\mathrm{DS}}$). For that, three samples with different $d_{\mathrm{WOx}}$ and one with a non-ferroelectric HfO$_2$ gate dielectric were realized.  $R_{\mathrm{DS}}$ was measured between source and drain after each \SI{2}{\micro\second} long write pulse ($V_{\mathrm{write}}$) applied to the gate (measurement scheme can be seen in Figure S4). For ease of comparison, $R_{\mathrm{DS}}$ is normalized by $R_{\mathrm{ON}}$ (Figure \ref{fig:WOx_thickness_HfO_PV_CV}c, d, e, f). A clear hysteresis in $R_{\mathrm{DS}}$ is observed for devices with a ferroelectric HZO gate dielectric. To confirm that the modulation of the channel resistance originates from $P_{\mathrm{r}}$ and not from another effect, an identical device with a non-ferroelectric HfO$_{2}$ gate dielectric was measured. Both have an \SI{8}{\nano\metre} thick WO$_{\mathrm{x}}$ channel. $R_{\mathrm{DS}}$ shows no hysteresis in the non-ferroelectric HfO$_{2}$ sample (Figure \ref{fig:WOx_thickness_HfO_PV_CV}c) and further proves that the hysteresis originates from the ferroelectricity in HZO.
In addition to the polarization in the HZO, the type and concentration of the free charge carriers\cite{Brotherton2013c,Bang2009} as well as $d_{\mathrm{WOx}}$ influence the on/off ratio. 
For a maximum reduction in the channel off-current, the polarization-field induced depletion width ($x_{\mathrm{d}}$) should be larger than $d_{\mathrm{WOx}}$. Using Poisson’s equation, the relationship between $x_{\mathrm{d}}$ and $N_D$ can be expressed as follows:\cite{Nakata2012,Bang2009,Goetzberger1966}

\begin{equation}\label{eq:depletionwidth}
x_{\mathrm{d}} = \frac{\epsilon_0 \epsilon_{\mathrm{WOx}}}{C_{\mathrm{HZO}}}\left[\left(1+\frac{ 2 C_{\mathrm{HZO}}^2 V_{\mathrm{GS}}}{q N_{\mathrm{D}} \epsilon_0 \epsilon_{\mathrm{WOx}}}\right)^{1/2} -1 \right],
\end{equation}
where $\epsilon_0$ is the vacuum permittivity, $\epsilon_{\mathrm{WOx}}$ the permittivity of  WO$_{\mathrm{x}}$ ($\epsilon_{\mathrm{WOx}} =  \num{189}$, see supplementary information), $C_{\mathrm{HZO}}$ is the HZO capacitance per unit area ( $C_{HZO}=\SI[per-mode=symbol]{3.14}{\micro\farad\per\centi\metre\squared}$, Figure S5b), and $V_{GS}$ is the polarization charge-induced potential across HZO. 
The carrier concentration ($N_{\mathrm{D}}=\SI{1.01E20}{\per\cubic\centi\metre}$), the channel resistivity ($\rho_{\mathrm{H}} = \SI{3.27E-01}{\ohm\centi\metre}$) and its mobility ($\mu_{\mathrm{H}}=\SI{0.19}{\square\centi\metre\per\volt\second}$) were determined by Hall measurements carried out on a similar sample. 
Using Eq. (\ref{eq:depletionwidth}), a depletion width $x_{\mathrm{d}} = $ \SIlist{1.7;3.3;4.8;6.4}{\nano\metre} for $V_{\mathrm{GS}} = $\SIlist{1;2;3;4}{\volt}, respectively was calculated (Figure S6b). For a constant polarization, the largest effect is obtained if $d_{\mathrm{WOx}}<x_{\mathrm{d}}=\SI{6.4}{\nano\metre}$ or $N_{\mathrm{D}}<\SI{1E20}{\per\cubic\centi\metre}$.
Three samples with different $d_{\mathrm{WOx}}$ were realized to benchmark this estimation with experimental data. BF-STEM measurements reveal $d_{\mathrm{WOx}}$ = \SIlist{8;11.3;15}{\nano\metre}, as reported in Figures \ref{fig:structural}b and S2a,b, respectively.
The polarization does not change between the three structures (Figures \ref{fig:WOx_thickness_HfO_PV_CV}a and S3a,b). By decreasing $d_{\mathrm{WOx}}$ from \SI{15}{\nano\metre}  to \SI{11.3}{\nano\metre}  and \SI{8}{\nano\metre}  the on/off ratio increases from \SI{\approx 1}{\percent} to \SI{\approx 5}{\percent} and \SI{\approx 90}{\percent}, respectively. Those results agree well with the $x_{\mathrm{d}}$ calculated by Eq. (\ref{eq:depletionwidth}). 

For neuromorphic applications multiple (analog) levels of the channel resistance, good retention properties, low device-to-device and cycle-to-cycle variations, fast updates, and low power consumption are important characteristics of ideal devices.\cite{Kim2017a,Gokmen2016,Gokmen,Yu2015} The exact requirements vary depending on the details of operation and from one implementation to the other. 
As an example, inference workloads would use off-line trained weights transferred to the chip to operate the network, and the precision of the weights (\num{>=3}bit) is more relaxed as in the case of a chip designed to perform on-line learning.\cite{Obradovic2018} 
In our device structure, weights are defined through the intermediate states of the channel resistance, enabled via the multi-domain nature of the ferroelectric HZO layer.\cite{Chanthbouala2012,Mulaosmanovic2017,Oh2017a}
\begin{figure*}[t!]
	\includegraphics[width=\textwidth]{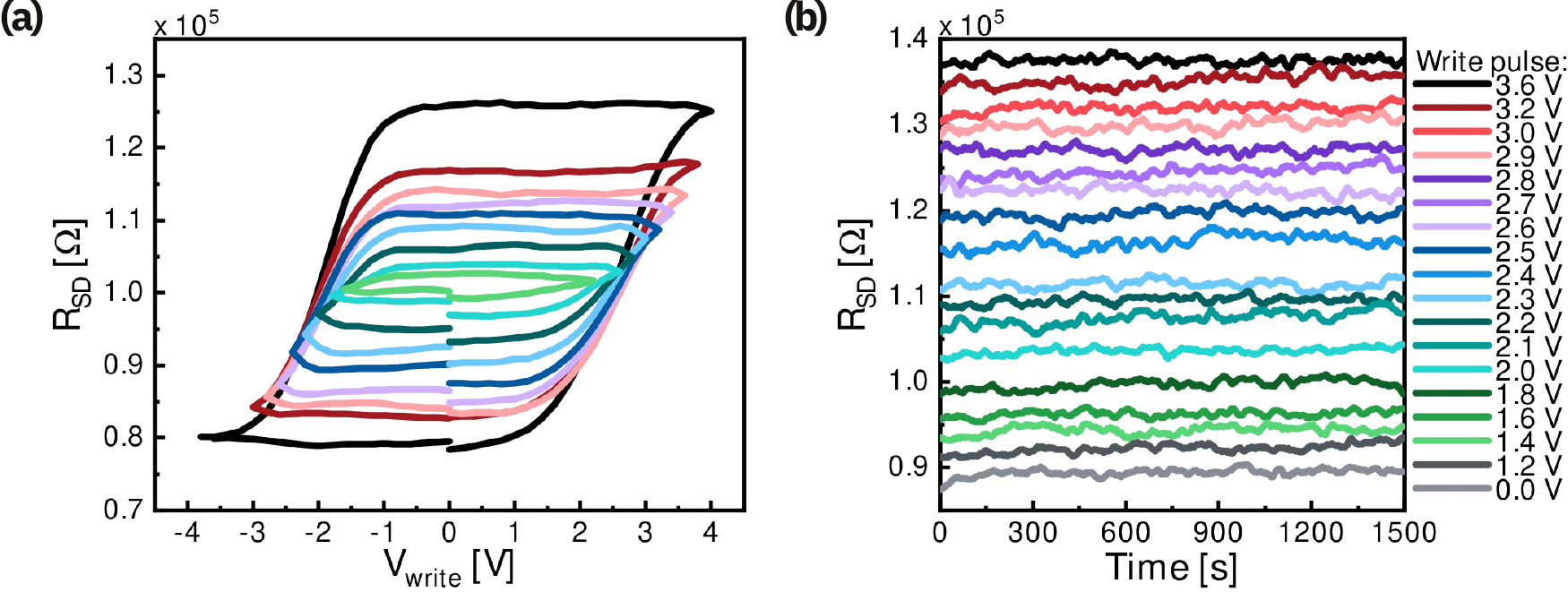}
  	\caption{Analog multi-level behavior of a FeFET of \SI{20}{\micro\metre} width and \SI{5}{\micro\metre} length. \textbf{(a)} The channel resistance ($R_{\mathrm{DS}}$) after the application of \SI{5}{\micro\second} write pulses ($V_{\mathrm{write}}$) of varying amplitudes. The different curves correspond to different consecutive measurements with reducing $V_{\mathrm{write}}$ range. \textbf{(b)} Retention measurement for \SI{1500}{\second} . $V_{\mathrm{read,D}} = \SI{200}{\milli\volt}$ was uninterruptedly applied while $R_{\mathrm{DS}}$ was measured every \SI{5}{\second}.}
  	\label{fig:MinorLoops_retention}
\end{figure*}
By switching only a subset of the domains, a state between $R_{\mathrm{ON}}$ and $R_{\mathrm{OFF}}$ can be set.\cite{Chanthbouala2012} The fraction of the switched ferroelectric domains depends on the amplitude, width, and number of the applied write pulses. Different pulsing schemes on HZO have been investigated in the past.\cite{Jerry2018} 
For on-line learning algorithms running on crossbar arrays integrated on CMOS, potentiation and depression pulse schemes with a constant pulse amplitude and width are preferred to those with varying amplitude.
Nevertheless, for the proof of concept the multi-state nature of a \SI{20}{\micro\metre} wide and \SI{5}{\micro\metre} long FeFET was investigated by applying voltage pulses of varying amplitudes, while keeping a fixed pulse duration of \SI{5}{\micro\second} (Figure \ref{fig:MinorLoops_retention}a). This pulse scheme results in the best linearity in potentiation and depression.\cite{Jerry2018}
By sweeping $V_{\mathrm{write}}$ from \SIrange{-4}{4}{\volt}, $R_{\mathrm{DS}}$ shows a hysteretic cycle from \SIrange{80}{125}{\kilo\ohm} with various intermediate states (on/off $\approx 1.55$).
By reducing the range of $V_{\mathrm{write}}$ numerous $R_{\mathrm{DS}}$ sub-loops can be accessed, as shown in Figure \ref{fig:MinorLoops_retention}a. The asymmetry in the hysteresis loop is due to the imprint in the ferroelectric layer. Furthermore, the retention properties have been studied, as demonstrated in Figure \ref{fig:MinorLoops_retention}b. First, an intermediate state was written by a \SI{5}{\micro\second} pulse. Then, a source-to-drain voltage $V_{\mathrm{DS}}= \SI{200}{\milli\volt}$ was applied for \SI{1500}{\second}, while $R_{\mathrm{DS}}$ was measured every \SI{5}{\second}. Between each measured intermediate state the FeFET was reset to its low resistive state ($R_{\mathrm{ON}}$) by setting $V_{\mathrm{write}} = \SI{-4}{\volt}$ during \SI{1}{\milli\second}. The FeFET showed stable retention properties for 18 differentiable channel resistances (>4bit) for the full \SI{1500}{\second}. The good retention measurement hints to an absence of depolarization or other screening mechanisms. The obtained multistate storage capability, the long retention and rather fast programming speed makes this FeFET suited for inference applications.

\begin{figure*}[t]
	\includegraphics[width=\textwidth]{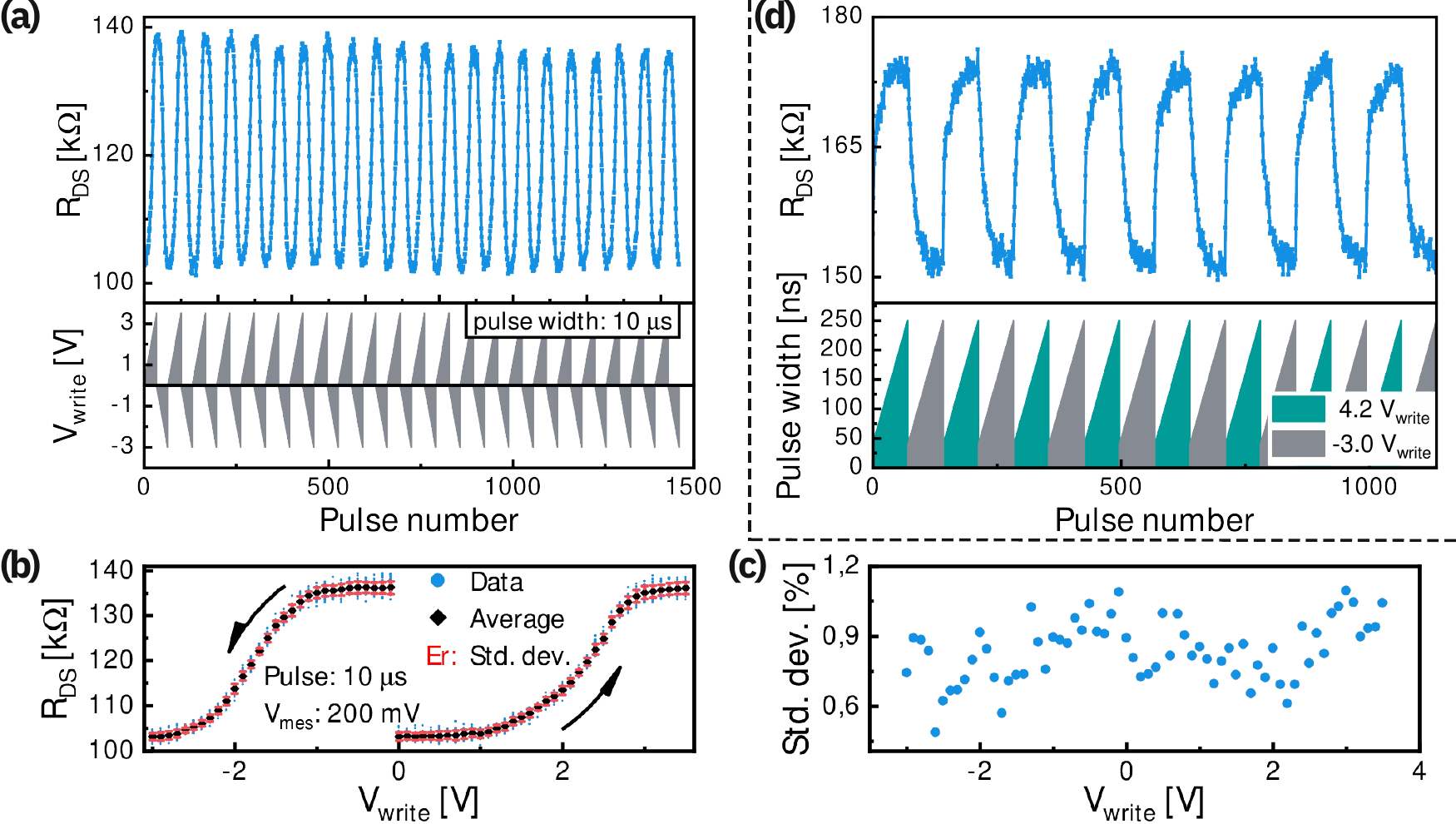}
  	\caption{Potentiation and depression of a \SI{20}{\micro\metre} wide and \SI{5}{\micro\metre} long FeFET. \textbf{(a)} The top panel shows multiple potentiation and depression cycles of the channel resistance ($R_{\mathrm{DS}}$) with varying pulse amplitude ($V_{\mathrm{write}}$) and constant pulse width ($t_{\mathrm{write}}$). The bottom panel shows the corresponding write pulse  sequence . After each pulse $R_{\mathrm{DS}}$ was measured. \textbf{(b)} Absolute cycle-to-cycle variation of $R_{\mathrm{DS}}$ showing the data, average and standard deviation error bars. \textbf{(c)} Standard deviation of the $R_{\mathrm{DS}}$ cycle-to-cycle variation in percent. \textbf{(d)}  Multiple potentiation and depression cycles of $R_{\mathrm{DS}}$ with increasing $t_{\mathrm{write}}$ from \SIrange{40}{250}{\nano\second} and constant $V_{\mathrm{write}}$. }
  	\label{fig:PotDep}
\end{figure*}
\begin{figure*}[t]
	\includegraphics[width=\textwidth]{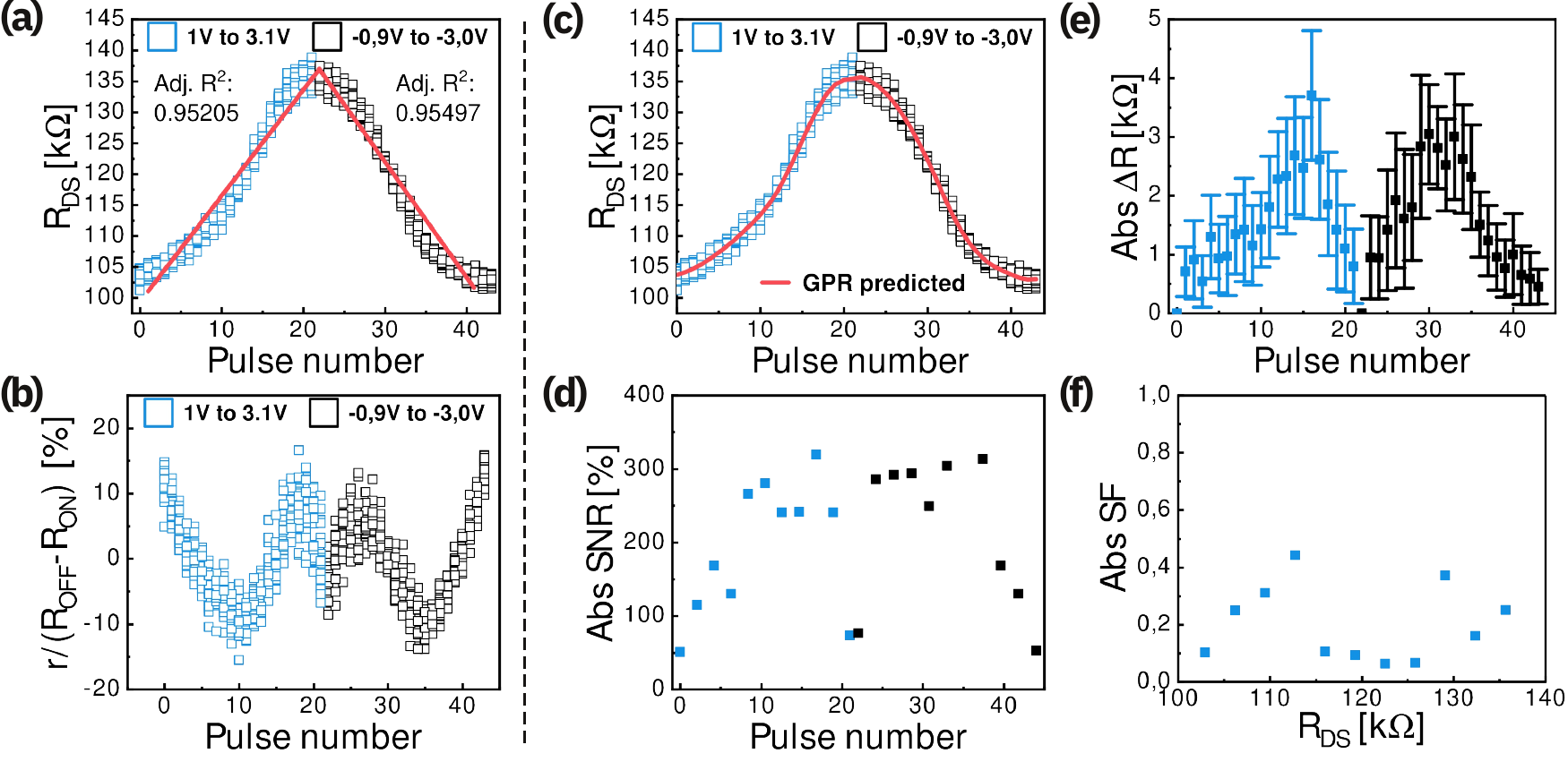}
  	\caption{Extraction of linearity and symmetry metric. Linear regression and the GPR methodology\cite{Gong2018} is applied to our FeFET data from multiple cycles with: 22 potentiation pulses (blue) with increasing amplitude (\SI{1}{\volt} to \SI{3.1}{\volt}) and 22 depression pulses (black) with decreasing negative amplitude (\SI{-0.9}{\volt} to \SI{-3}{\volt}). \textbf{(a)} $R_{\mathrm{DS}}$ as a function of pulse number and the linear regression fit (red). \textbf{(b)} Absolute residuals $r$ normalized by the channel resistance window.\textbf{(c)} Channel resistance ($R_{\mathrm{DS}}$) as a function of pulse number and the GPR predicted noise free signal (red). \textbf{(d)} Absolute SNR for each potentiation and depression pulse. \textbf{(e)} Absolute change of $R_{\mathrm{DS}}$ after each potentiation and depression pulse. \textbf{(f)} Symmetry factor (SF) as a function of $R_{\mathrm{DS}}$.}
  	\label{fig:Symmetry_linearity}
\end{figure*}

For on-chip learning, artificial synapses require a finer mesh of intermediate levels. In addition, symmetric and linear potentiation and depression are desirable. With respect to symmetry the field-driven ferroelectric switching is advantageous to other technologies that often show abrupt or unidirectional switching.\cite{Yu2015,Jerry2018} The requirement of low variability is relaxed as the training occurs on a specific hardware and thus incorporates the variability in its solution.\cite{Obradovic2018} To investigate the linearity and symmetry of the potentiation and depression, multiple write pulses of increasing and decreasing amplitude were applied. For the potentiation $V_{\mathrm{write}}$ was increased from \SIrange{0}{3.5}{\volt} and for the depression decreased from  \SIrange{0}{-3}{\volt} with \SI{100}{\milli\volt} steps (Figure \ref{fig:PotDep}a). The duration of the write pulses was kept constant at \SI{10}{\micro\second}. 
When averaging over several cycles (Figure \ref{fig:PotDep}b), multiple states with small standard deviation are observed. Normalizing the cycle-to-cycle standard deviation by $R_{\mathrm{ON}}$ reveals a constant value of about \SI{1}{\percent} (Figure \ref{fig:PotDep}c). The number and overlap of states are defined by the potentiation and depression step size. The latter could be reduced further to increase the resolution.
When fitting the potentiation range from \SIrange{1}{3.1}{\volt} and depression range from \SIrange{-0.9}{-3.0}{\volt}  by linear regression (Figure \ref{fig:Symmetry_linearity}a), an adjusted residual-square value of 0.952 is obtained. The residuals normalized by the $R_{\mathrm{DS}}$ window as a function of pulse number is depicted in Figure \ref{fig:Symmetry_linearity}b. 
For a more detailed analysis of the symmetry, Gaussian process regression (GPR) was used to predict a noise free signal (Figure \ref{fig:Symmetry_linearity}c)\cite{Gong2018}.
Plotting $\Delta R$ (Figure \ref{fig:Symmetry_linearity}e) and the signal to noise ratio (SNR, Figure \ref{fig:Symmetry_linearity}d) as a function of pulse number reveals diminishing $\Delta R$ and noisier signals towards the extremes.
The symmetry factor (SF) was then calculated using the following equation:\cite{Gong2018} 

\begin{equation}\label{eq:SF}
SF = \lvert\frac{\Delta R_+ - \Delta R_- }{\Delta R_+ + \Delta R_-} \rvert,
\end{equation}
where $\Delta R_+$ is the potentiation and $\Delta R_-$ is the depression change in resistance at a certain resistance level. By this definition, SF can take values between 0  and 1 where 0 is the perfect symmetry.
The less linear the range of the data becomes, the larger is SF (Figure \ref{fig:Symmetry_linearity}d). The average across the full resistance range is $SF=0.20$ while the most linear part in the center reaches a very good symmetry factor of $SF=0.08$.

Short programming pulses are advantageous as fast writing and low-power consumption are important for neuromorphic applications. 
By varying the pulse width from \SIrange{40}{250}{\nano\second} with a fixed amplitude (Figure \ref{fig:PotDep}d), already the shortest applied pulse of \SI{40}{\nano\second} (equipment limit) changes the resistance and demonstrates very fast writing capabilities of the FeFET.
It is expected that even shorter pulses could successfully program the device.\cite{Chanthbouala2012} 
In our device, little energy is consumed while writing a state. When applying $V_{\mathrm{write}}=\SI{3.5}{\volt}$ a gate current of $I_{\mathrm{gate}}=\SI{3.02E-8}{\ampere}$ is measured. Applying a write pulse duration of $t_{\mathrm{write}}=\SI{200}{\nano\second}$ results in $E=\frac{V_{\mathrm{write}}\cdot I_{\mathrm{gate}} \cdot t_{\mathrm{write}}}{w \cdot l}=\SI{2.1E-17}{\joule\per\micro\metre\squared}$, where $l$ is the length and $w$ the width of the gate.

\section{Conclusion}

We propose a device concept based on the ferroelectric field effect into a thin WO$_{\mathrm{x}}$ channel using HZO gate dielectric, that can be used as a synaptic element in hardware-supported neural networks. The fabrication process is compatible with the integration in the Back End Of Line of CMOS technology and is using earth-abundant materials, which is making it attractive for large-scale integration. By comparing HZO and HfO$_2$ based devices, and carefully analyzing capacitor and transistor data, we unambiguously show that the channel resistance is directly coupled to the polarization of the HZO layer and can be programmed in a non-volatile manner. Multilevel states programmed over more than 4-bits depth with a good retention and an almost symmetric potentiation and depression is obtained, together with a low programming energy. The property of the WO$_{\mathrm{x}}$ layer and the geometry of the device can be arranged so that a well-suited resistance range is obtained, favorable to build large scale arrays. The proposed device exhibit therefore promising metrics when considered as a synaptic element for processing cores supporting artificial neural networks. Future work will concentrate on controlling the channel thickness and the carrier concentration of WO$_{\mathrm{x}}$  to increase the on/off ratio, so that the device can  be operated strictly in the linear region, without ever fully switching all the domains to the same polarization. This is expected to improve symmetry and to allow a constant pulse scheme for potentiation and depression, which is more friendly to learning algorithms.

\section{Experimental}
\indent
\textbf{Sample preparation.} 
Our FeFET is a bottom/gate device with shared gate. The gate contact is accessed through the Si $n^+$ substrate. First, \SI{10}{\nano\metre}  TiN was deposited using a tetrakis(dimethylamino)titanium (TDMAT) precursor and N$_2$/H$_2$ plasma in an Oxford Instruments plasma enhanced atomic layer deposition (PEALD) system. An approximately \SI{10}{\nano\metre} thick layer of HZO was grown in a process using alternating cycles of tetrakis(ethylmethylamino)hafnium (TEMAH), and ZrCMMM ((MeCp)2Zr(OMe)(Me)) at \SI{300}{\celsius}. Rutherford Back Scattering (RBS) analysis of the film (not shown) indicated an actual film composition of Hf$_{0.57}$Zr$_{0.43}$O$_2$. The sample was then immediately transferred to a sputter chamber for the deposition of \SI{4}{\nano\metre}  W. For the crystallization of HZO a millisecond flash lamp anneal (ms-FLA)\citep{OConnor2018} with a background temperature of \SI{375}{\celsius} was performed. After crystallization the \SI{4}{\nano\metre}  W was reduced to \SI{\approx 2.5}{\nano\metre}  by Ar sputtering. The W was then crystallized and oxidized to \SI{10}{\nano\metre} WO$_3$ in a rapid thermal annealer (RTA) at \SI{350}{\celsius} for \SI{6}{\minute}  with  \SI{50}{\sccm} O$_2$. Afterwards a reduction of the WO$_3$ to WO$_{\mathrm{x}}$ was performed in a RTA by H$_2$ annealing at \SI{150}{\celsius} and vacuum annealing at \SI{350}{\celsius}. WO$_{\mathrm{x}}$ was further thinned by $Ar$ sputtering to \SI{8}{\nano\metre}. Source and drain were deposited by sputtering and liftoff. The passivation consists of \SI{5}{\nano\metre} Al$_2$O$_3$ by thermal ALD (precursor) and \SI{100}{\nano\metre} SiO$_2$ by plasma-enhanced chemical vapor deposition (PECVD). Vias were etched using a reactive ion etcher (RIE) with a CHF$_3$/O$_2$ plasma. Finally, the contacts were realized by depositing \SI{100}{\nano\metre} W by sputtering and defined in an RIE with a SF$_6$/O$_2$ plasma.

\textbf{Structural Characterization.} 
Grazing incidence X-ray diffraction (GIXRD) measurements were performed in a Bruker D8 Discover diffractometer equipped with a rotating anode generator. TEM lamellas have been prepared by Focused Ion Beam using a FEI FIB Helios FEI Helios NanoLab 450S and investigated with a double spherical aberration-corrected JEOL JEM-ARM200F microscope. Bright field STEM (BF-STEM) images have been acquired at 200 kV and Energy Dispersive x-ray Spectroscopy (EDS) line profiles have been performed using a liquid-nitrogen-free silicon drift detector.

\textbf{Electrical Characterization.}
$R_{\mathrm{DS}}-V_{\mathrm{write}}$ and retention were measured using an Agilent B1500. $V_{\mathrm{write}}$ pulses were generated by a WGFMU and RSU module for the Agilent B1500 and applied to source and drain simultaneously while grounding the gate (Figure S4a). $R_{\mathrm{DS}}$ was measured by applying an IV-sweep from \SIrange{-200}{200}{\milli\volt} to the drain while having the source connected to ground (Figure S4b). $R_{\mathrm{DS}}$ was then determined by averaging the resistance at \SI{\pm 200}{\milli\volt}. 
$P-V$ loops on HZO were recorded using a TF Analyzer 2000 from AixAct. The signal of \SI{5}{\kilo\hertz} was applied to the top W/WOx contact while the bottom TiN/$n^+$ Si contact (substrate) was grounded. For the wake-up of HZO, \num{e5} cycles of \SI{\pm 3.8}{\volt} and \SI{100}{\kilo\hertz} were applied.

\begin{suppinfo}

The Supporting Information is available free of charge on the ACS Publications website at DOI:
\begin{itemize}
  	\item Additional data concerning the endurance of MFM and MSFM structures, BF-STEM and $P-V$ measurements on additional samples, electrical measuring schemes, capacitance measurements and permittivity and depletion width calculations. (PDF)
  		
\end{itemize}

\end{suppinfo}

\begin{acknowledgement}
We acknowledge helpful discussions with Nanbo Gong and Takashi Ando.	 
This project has received funding from the European Commission under grant agreement H2020-ICT-2016-1-732642 (ULPEC).

\end{acknowledgement}

\bibliography{bibtex/PhD-FeFET-Paper.bib}

\end{document}


\newpage

\begin{center}
	\includegraphics[width=\textwidth]{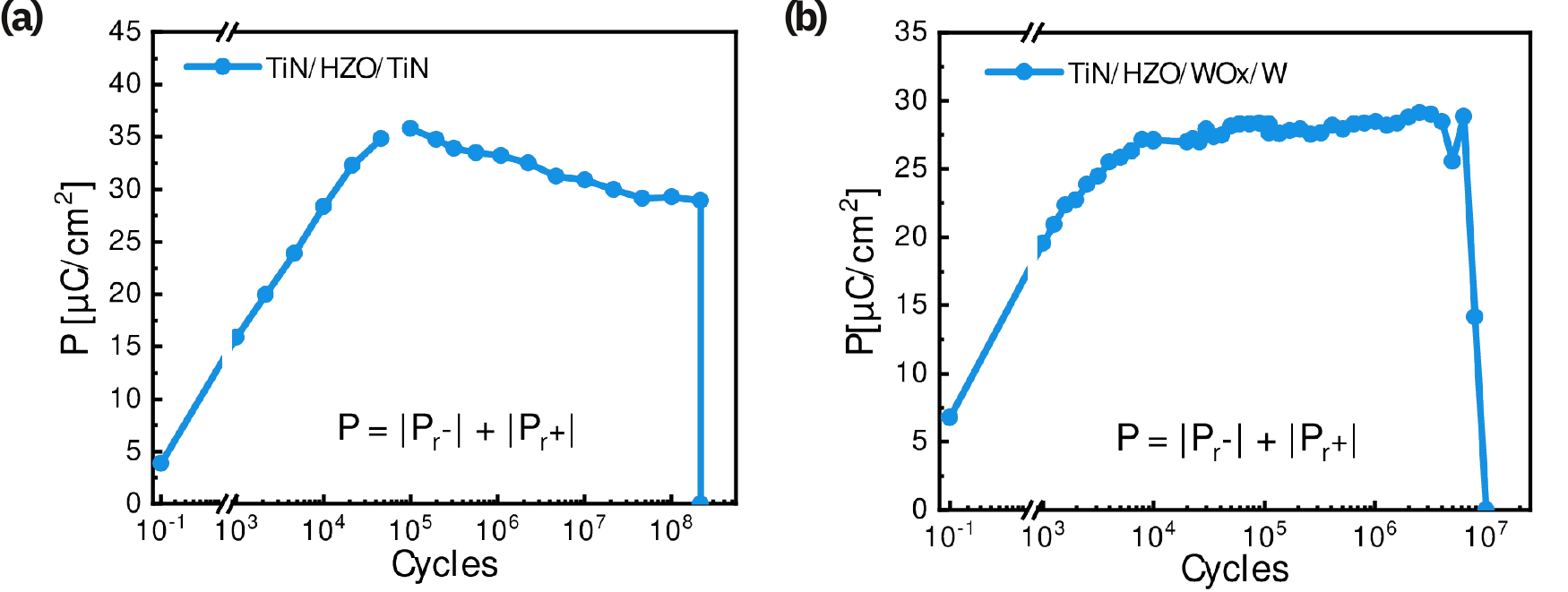}
  	\captionof{figure}{Endurance measurements of \SI{10}{\nano\metre} HZO. The total remanent polarization \mbox{($P=\lvert P_{\mathrm{r-}} \rvert + \lvert P_{\mathrm{r+}} \rvert$)} was determined by positive up negative down (PUND) measurements with \SI{1}{\kilo\hertz} and \SI{\pm 3.5}{\volt}. The cycling frequency was set to \SI{1}{\kilo\hertz} up to \num{e4} cycles, \SI{10}{\kilo\hertz} up to \num{e5} cycles and \SI{100}{\kilo\hertz} for cycles above \num{e5}: \textbf{(a)} The TiN/HZO/TiN MFM configuration was cycled at \SI{\pm 3.5}{\volt}. \textbf{(b)} The W/WO$_{\mathrm{x}}$/HZO/TiN MSFM configuration was cycled at \SI{\pm 3.0}{\volt} with a \SI{-0.5}{\volt} offset.}
  	\label{Supfig:endurance}
\end{center}

\vspace{1in}

\begin{center}
	\includegraphics[width=\textwidth]{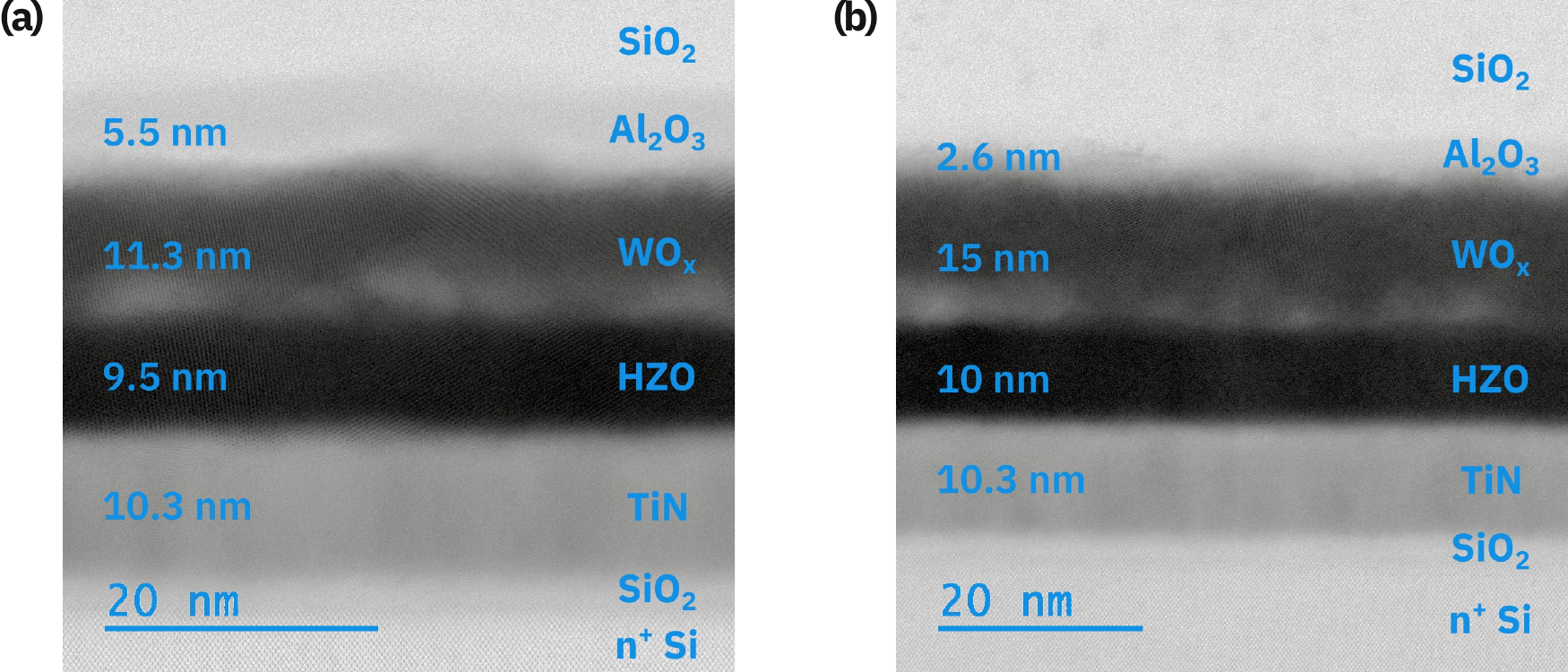}
  	\captionof{figure}{Cross-sectional BF-STEM images of the samples from the WO$_{\mathrm{x}}$ thickness series: \textbf{(a)} $d_{\mathrm{WOx}} = \SI{11.3}{\nano\metre}$, \textbf{(b)} $d_{\mathrm{WOx}} = \SI{15}{\nano\metre}$}
  	\label{Supfig:STEM}
\end{center}
\newpage

\begin{center}
	\includegraphics[width=\textwidth]{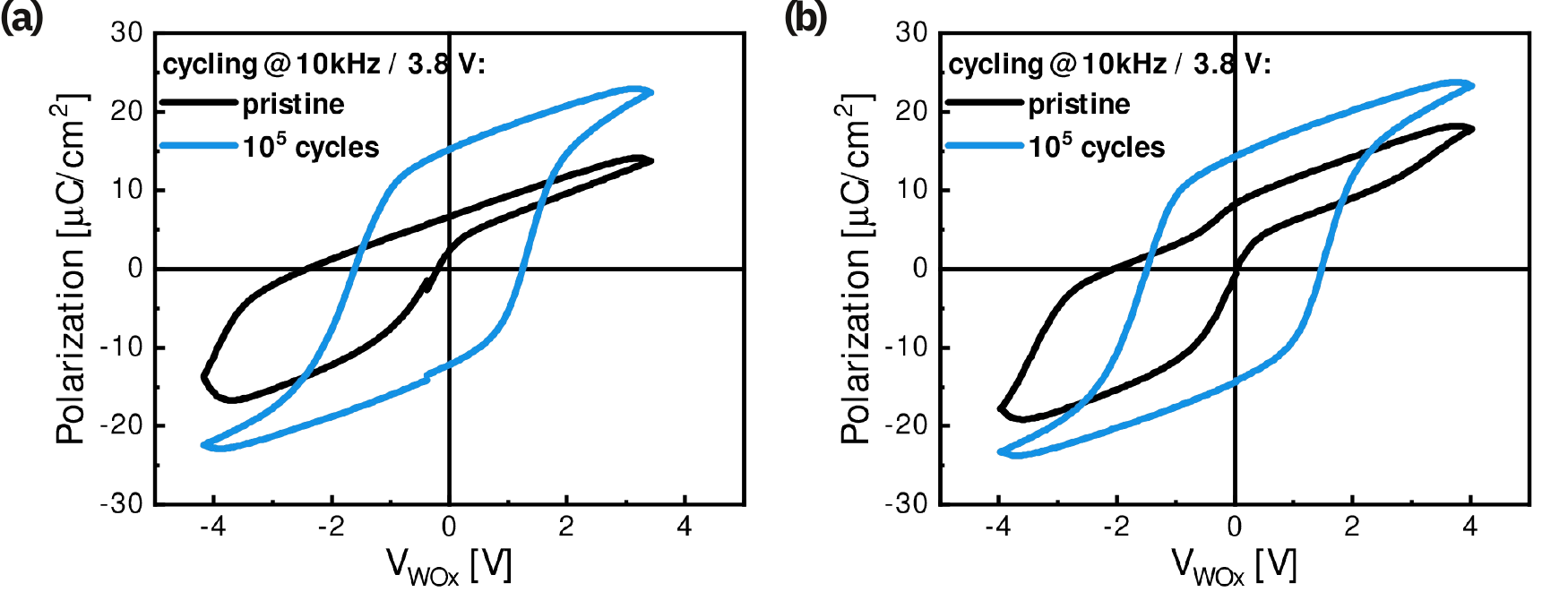}
  	\captionof{figure}{P-V measurements on the samples from the WO$_{\mathrm{x}}$ thickness series. Polarization versus voltage ($P-V$) characteristics measured on \SI{60 x 60}{\micro\metre} W/WO$_{\mathrm{x}}$/HZO/TiN/n$^{\mathrm{+}}$Si MFM structures at \SI{5}{\kilo\hertz} in the pristine state and after \num{e5} cycles: \textbf{(a)} $d_{\mathrm{WOx}} = \SI{11.3}{\nano\metre}$, \textbf{(b)} $d_{\mathrm{WOx}} = \SI{15}{\nano\metre}$.}
  	\label{Supfig:PE}
\end{center}
\vspace{1in}

\begin{center}
	\includegraphics[width=\textwidth]{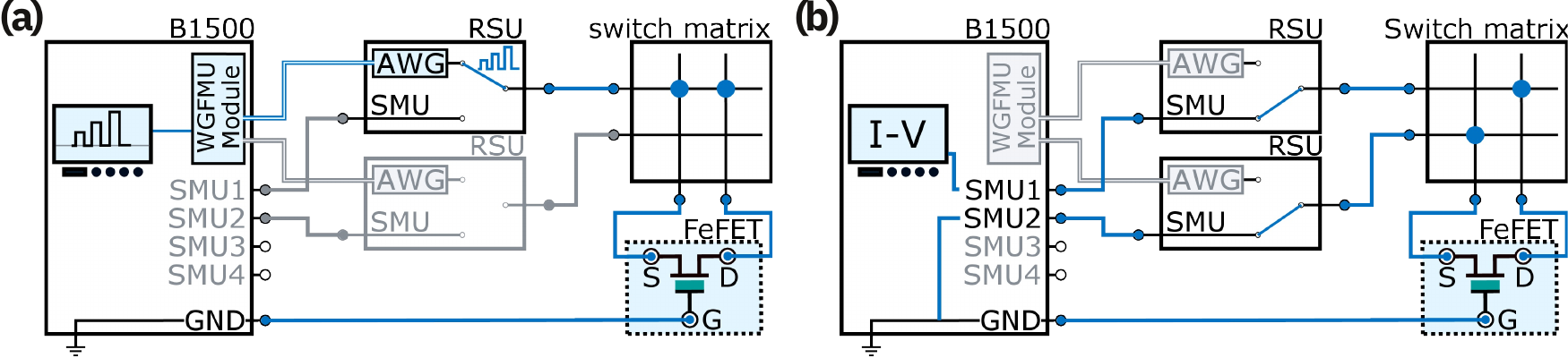}
  	\captionof{figure}{Write and read schematic showing a semiconductor parameter analizer (B1500) with the waveform generator/fast measurement unit (WGFMU) and its two remote-sense and switch units (RSU): A state is written by applying a pulse to source (S) and drain (D) while the gate (G) is grounded. The channel resistance is read by applying an IV-sweep from \SIrange{-200}{200}{\milli\volt} to D while having S connected to the common ground through SMU2. }
  	\label{Supfig:write_read}
\end{center}

\newpage

\begin{center}
	\includegraphics[width=\textwidth]{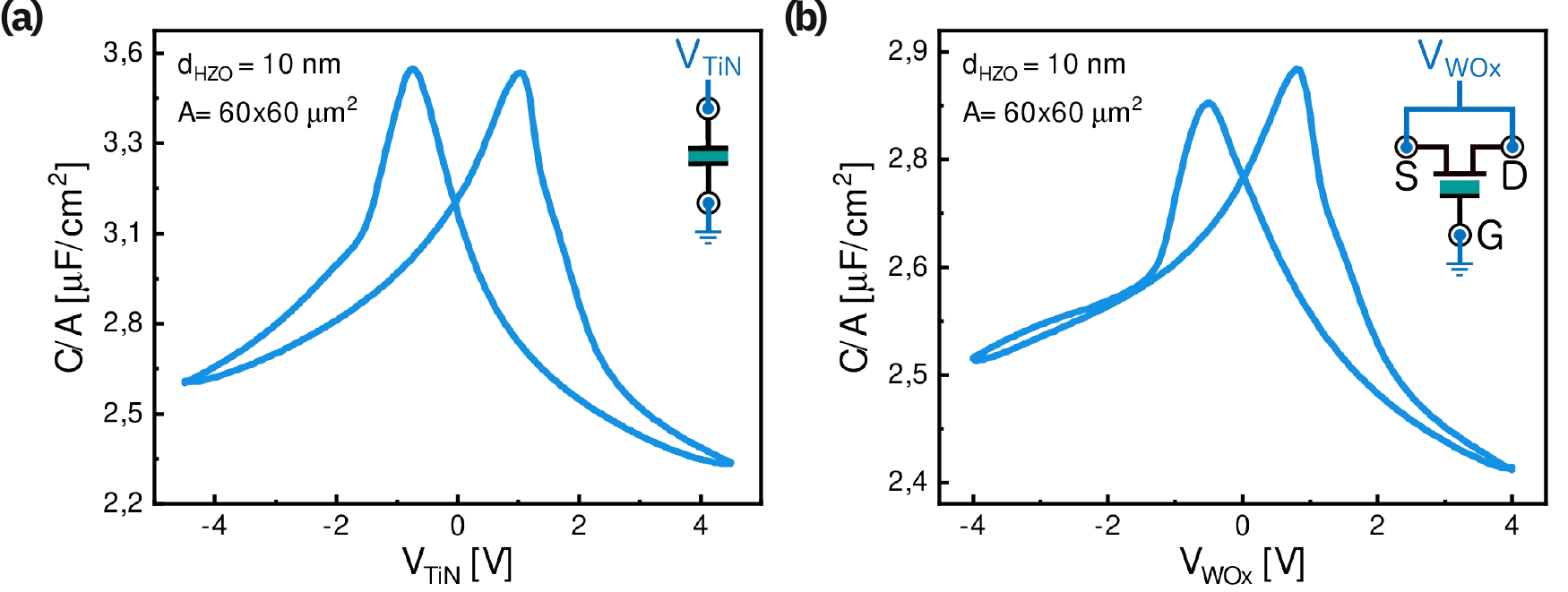}
  	\captionof{figure}{Capacitance measurements on a  \SI{60 x 60}{\micro\metre} \textbf{(a)} TiN/WO$_{\mathrm{x}}$/TiN MFM and \textbf{(b)} W/WO$_{\mathrm{x}}$/HZO/TiN/n$^{\mathrm{+}}$Si MSFM structure}
  	\label{Supfig:CV}
\end{center}
\vspace{1in}

\subsection*{WO$_{\mathrm{x}}$ permitivity}
The permittivity of WO$_{\mathrm{x}}$ ($\epsilon_{\mathrm{WOx}}=189$) was calculated using the following equation of two capacitances in series: 
\begin{equation}\label{eq:capacitance}
\frac{1}{C_{\mathrm{WOxHZO}}} =\frac{1}{C_{\mathrm{HZO}}} + \frac{d_{\mathrm{WOx}}}{\epsilon_0*\epsilon_{\mathrm{WOx}}*A},
\end{equation}
where $C_{\mathrm{WOxHZO}}$ is the capacitance of the W/WO$_{\mathrm{x}}$/HZO/TiN stack, $C_{\mathrm{HZO}}$ the capacitance of TiN/HZO/TiN stack, $d_{\mathrm{WOx}} = \SI{8}{\nano\metre}$ the thickness of the WO$_{\mathrm{x}}$ channel, $\epsilon_0$ the vacuum permittivity and $A = \SI{3600}{\square\micro\metre}$ the area of the capacitor. From Figure \ref{Supfig:CV}a we get $C_{\mathrm{HZO}}=\SI{1.13E-10}{\farad}$ and from Figure \ref{Supfig:CV}b we get $C_{\mathrm{WOxHZO}}=\SI{9.9E-11}{\farad}$.

\newpage

\begin{center}
	\includegraphics[width=0.5\textwidth]{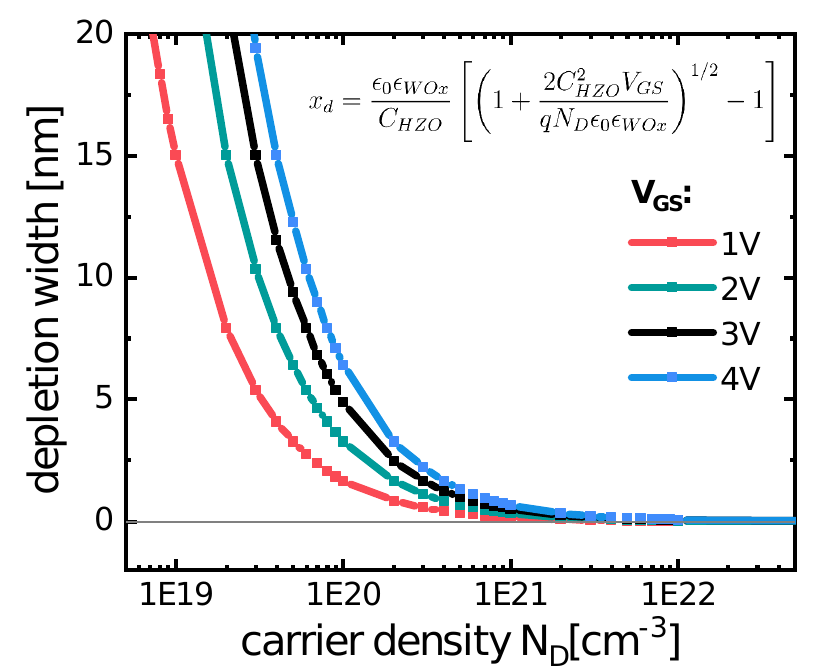}
  	\captionof{figure}{Depletion width as a function of carrier concentration N$_{\mathrm{D}}$ calculated for different V$_{\mathrm{GS}}$}
  	\label{Supfig:Hall}
\end{center}